\documentclass[preprint,showpacs,preprintnumbers,endfloats*,amsmath,amssymb,linenumbers,prb]{revtex4-1}
\usepackage{graphicx}% Include figure files
\usepackage{dcolumn}% Align table columns on decimal point
\usepackage{bm}% bold math
\newcommand{\vecvar}[1]{\mbox{\boldmath$#1$}}
\newcommand{\e}{\mbox{e}}

\begin{document}

\preprint{PRESAT-8502}

\title{First-principles calculation of scattering potentials of Si-Ge and Sn-Ge dimers on Ge(001) surfaces}

\author{Tomoya Ono}
\affiliation{Graduate School of Engineering, Osaka University, Suita, Osaka 565-0871, Japan}

\date{\today}% It is always \today, today,

\begin{abstract}
The scattering potential of the defects on Ge(001) surfaces is investigated by first-principles methods. The standing wave in the spatial map of the local density of states obtained by wave function matching is compared to the image of the differential conductance measured by scanning tunneling spectroscopy. The period of the standing wave and its phase shift agree with those in the experiment. It is found that the scattering potential becomes a barrier when the electronegativity of the upper atom of the dimer is larger than that of the lower atom, while it acts as a well in the opposite case.
\end{abstract}

\pacs{73.20.Hb, 72.10.Fk, 82.20.Wt}% PACS, the Physics and Astronomy
\maketitle
\section{Introduction}
\label{sec:introduction}
With the progress of new techniques for the atomic-scale manipulation and modification of materials, there is considerable interest in the electron scattering properties of nanostructures both experimentally and theoretically. Scanning tunneling microscopy (STM) \cite{stm} and mechanically controllable break junctions \cite{ruitenbeek} have been extensively used to study the conductance of atomic-scale systems suspended between two electrodes so far, and today it is already well known that the conductance of a single strand of atomic wire is quantized in the unit of $2e^2/h$, where $e$ is the electron charge and $h$ is Planck's constant.\cite{rubio-takayanagi-kizuka} Although these techniques are powerful tools to understand electron transport phenomena of such minute systems, it has not been easy to discuss the contribution of local chemical bonds to the electron scattering because these techniques only measure the current between two electrodes with an applied bias voltage. On the other hand, the spatial maps of the local density of states (LDOS) obtained by scanning tunneling spectroscopy (STS) can provide the images of standing waves, which give important information about the dispersion relation of surface states \cite{eigler-avouris,li-chen} as well as about the electron scattering process at the potential barrier.\cite{eigler-avouris}

Tomatsu {\it et al.} demonstrated the standing wave around oppositely buckled Ge dimers on a Ge(001) surface and discussed the difference in scattering potential between the Si-Ge and Sn-Ge dimers.\cite{tomatsu0,tomatsu1} They also carried out first-principles calculations based on the density functional theory (DFT) \cite{dft} using periodic supercells to identify the surface atomic and electronic structures observed in STM images and STS spectra. However, the standing waves emerge as a result of the difference in the coefficients between incident and reflected waves of scattering wave functions, which cannot be calculated by models under periodic boundary conditions along the scattering direction. Thus, it is not straightforward to determine the scattering potential using the periodic models.

The scattering properties of atomic-scale structures are studied by DFT calculations combined with nonequilibrium Green's functions \cite{keldysh} or wave function matching methods.\cite{ando,obm,icp} In these approaches, the scatterers are sandwiched between electrodes that extended semi-infinitely and are connected to bulks; thus, wave functions extending over the entire system can be correctly described as a scattering state distributed as a result of the existence of scatterers when an electron comes from infinitely deep inside the electrode. Although these approaches have been used to investigate the transport properties of atoms, molecules, and thin films suspended by two electrodes so as to model the experiments using the STM and mechanically controllable break junctions, to the best of my knowledge, no studies have attempted to obtain an understanding of the relationship between local chemical bonds and their scattering properties by examining the scattering potential in first-principles transport calculations.

In this study, I apply the first-principles transport calculation to the investigation of the standing waves due to scattering at Si-Ge and Sn-Ge dimers on a Ge(001) surface. The real-space procedure of the DFT calculation combined with the overbridging-boundary matching (OBM) method provides the reflection and transmission coefficients of the incident electrons, which enable us to determine the scattering potential and the phase shift of the standing waves. The calculated phase shift of the standing waves is in good agreement with that measured in experiments.\cite{tomatsu1} The defects act as a barrier when a Si (Sn) atom exists at the lower (upper) position of the dimer, while they behave as a well in the case of the other way around. It is found that the scattering potential is related to the stabilization of the $\pi$ bands of the Ge(001) surface due to the Jahn-Teller effect of the dimers and the difference in electronegativity between Ge and the impurity atom.

The first-principles calculation method used here to obtain the electron scattering properties is based on the real-space finite-difference approach,\cite{chelikowsky,icp,tsdg} which enables us to determine the self-consistent electronic ground state with a high degree of accuracy by a timesaving double-grid technique.\cite{icp,tsdg} Moreover, the real-space calculations eliminate the serious drawbacks of the conventional plane-wave approach, such as its inability to describe nonperiodic systems accurately. The scattering properties of Ge-Si and Ge-Sn dimers on a Ge(001) surface suspended between Ge(001)-(2$\times$2) surfaces are examined by making use of this advantage of the combination of the real-space finite-difference approach and the OBM method.\cite{obm,icp}

The rest of this paper is organized as follows. In Sec. II, I briefly describe the computational methods and models used in this study. My results are presented and discussed in Sec. III. Finally, I summarize my findings in Sec. IV.

\section{Computational methods and models}
\begin{figure}
\begin{center}
\includegraphics{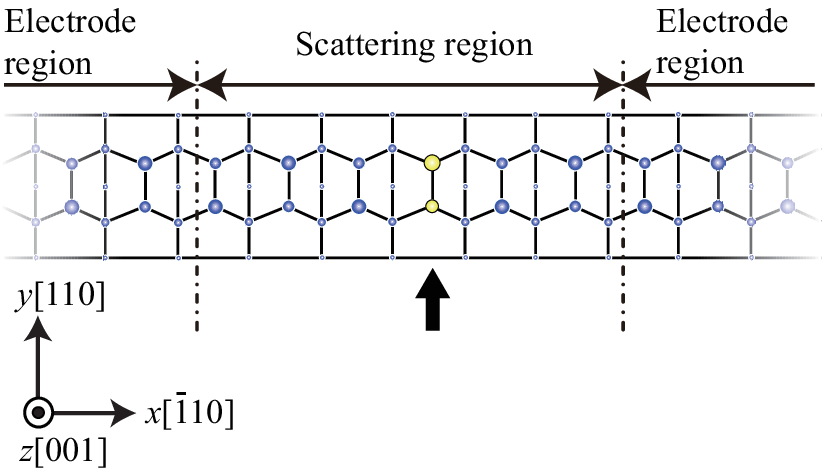}
\caption{Schematic image of computational model. Dark (blue) spheres are Ge atoms and light (yellow) spheres represent the dimer replaced by a Ge-Si or Ge-Sn dimer, which is indicated by the arrow. Atoms are denoted by large and small spheres according to the distance from the surface. \label{fig:1}}
\end{center}
\end{figure}

Figure~\ref{fig:1} shows the computational model, where the scattering region is connected to the Ge(001)-(2$\times$2) surfaces. The scattering region consists of six Ge atom layers with a (2$\times$6) lateral cell, and the topmost atoms are buckled so as to form a surface reconstructed structure, while the bottommost layers are terminated by hydrogen atoms. The dimensions of the cells are $L_x=23.92$ \AA, $L_y=7.97$ \AA, and $L_z=17.94$ \AA \hspace{2mm} for the scattering region, and $L_x=7.97$ \AA, $L_y=7.97$ \AA, and $L_z=17.94$ \AA \hspace{2mm} for the Ge(001)-(2$\times$2) surface electrode region, where $L_x$ and $L_y$ are the lengths in the $x$- and $y$-directions parallel to the surface, respectively, and $L_z$ is the length in the $z$-direction. The Ge-Ge dimer at the center of the scattering region is replaced by Ge-Si or Ge-Sn dimer as a defect. When the Si and Ge atoms are located at the lower and upper sides of the dimer, respectively, the dimer is referred as an SiL dimer. Other dimers are named in a similar manner. During the optimization of the atomic configurations in both the electrode and scattering regions, all atoms except for those in the bottommost layers are relaxed until all the force components drop below 0.05 eV/\AA \hspace{2mm} by imposing the periodic boundary condition in the $x$-, $y$-, and $z$-directions. The grid spacing is set at $\sim$ 0.20 \AA \hspace{2mm} and {\it k}-space integrations are performed with $4 \times 4\vecvar{k}$ points in the irreducible wedge of a two-dimensional Brillouin zone of the Ge(001)-(2$\times$2) surface. Exchange correlation effects are treated by a local density approximation \cite{lda} of the DFT, and the projector augmented wave method \cite{paw} is used to describe the electron-ion interaction.

In the OBM calculation performed to obtain the scattering wave functions, the norm-conserving pseudopotentials \cite{norm} of Troullier and Martins \cite{tmpp} are employed instead of the projector augmented wave method. To determine the Kohn-Sham effective potential, a supercell is used under a periodic boundary condition, and then the scattering wave functions are computed under the semi-infinite boundary condition obtained non-self-consistently. It has been reported that this procedure is just as accurate in the linear response regime but significantly more efficient than performing computations self-consistently on a scattering-wave basis.\cite{kong2}

\section{Results and discussion}
\label{sec:Results and discussion}
\begin{figure*}
\begin{center}
\includegraphics{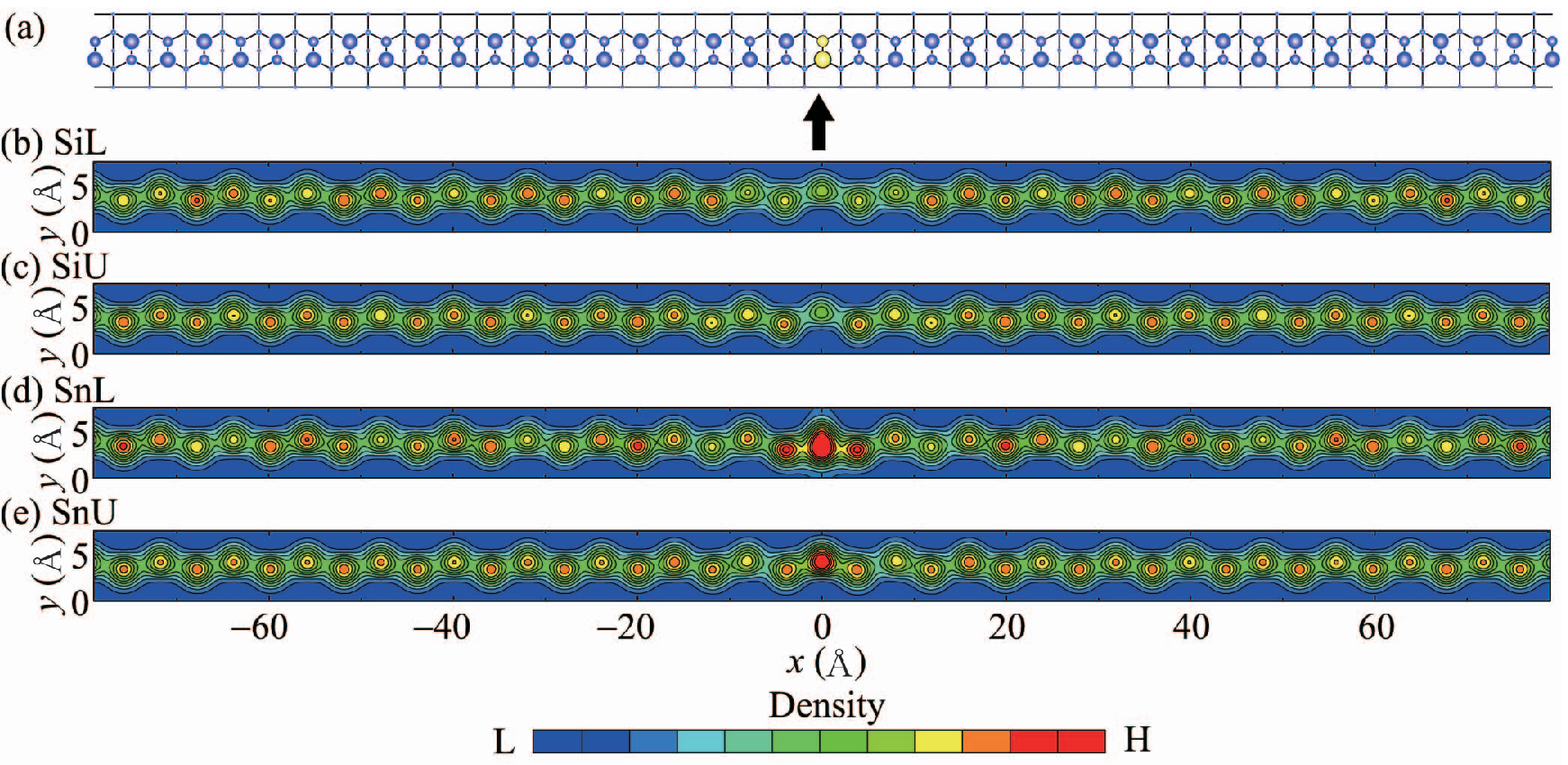}
\caption{(Color) Surface atomic structure (a) and spatial images of standing waves for (b) SiL, (c) SiU, (d) SnL, and (e) SnU dimers. In (a), the meanings of symbols are the same as those in Fig.~\ref{fig:1}. In (b), (c), (d), and (e), each contour represents a density of $0.496 \times 10^{-5}$ {\it e}/\AA$^3$/eV/spin higher or lower than that of the adjacent contours. \label{fig:2}}
\end{center}
\end{figure*}

Figure~\ref{fig:2} shows the charge density distribution of the scattering wave functions for the electrons propagating from both the left and right electrodes with an energy of $E_F+0.55$ eV, which corresponds to the spatial image of the differential conductance in the STS spectrum.\cite{tomatsu1} Here, $E_F$ is the Fermi energy. The standing waves on the $xy$ plane, which is 2.7 \AA \hspace{2mm} from the upper Ge atom of the Ge(001) surface, are plotted. It is found that the standing wave emerges in all the models. Since only $\pi^*$ band of the Ge(001)-(2$\times$2) surface is the propagating wave at just above the Fermi level, the scattering wave functions $\psi$ in the incident electrode is written as $\psi(\vecvar{r})=\e^{ik_xx}u(\vecvar{r})+R\e^{-ik_xx}u^*(\vecvar{r})$ with $u(\vecvar{r})$ and $R$ being the periodic part of the wave functions of the electrodes and reflection coefficient, respectively, and its charge density is expressed as $[1+|R|^2+2 \mbox{Re}(R) \cos(2k_xx)]u^*(\vecvar{r})u(\vecvar{r})$. In the present case, the $\pi^*$ band crosses the energy of $E_F+0.55$ eV at $k_x=0.454 \times \pi/L_x$ and the period of the standing waves is $\pi$/2$k_x$ (=17.6 \AA), which is slightly shorter than that observed in the experiment ($\sim 22$ \AA). As the decrease of the energy, the crossing points of $k_x$ between the $\pi^*$ band and the Fermi energy becomes lower,\cite{band} resulting in the longer period of the standing wave. Indeed, I have confirmed that the period of the standing wave with the energy of $E_F+0.45$ eV is 20.3 \AA. The period of the standing waves approaches to that observed in the experiment when the $\pi^*$ band is shifted rigidly so as to compensate the underestimation of the band gap due to the local density approximation. In addition, taking into account the fact that the period decreases as the sample bias increases in the spatial image of the STS spectrum in Ref.~\onlinecite{tomatsu1}, these results agree with those of the experiments.

\begin{figure*}
\begin{center}
\includegraphics{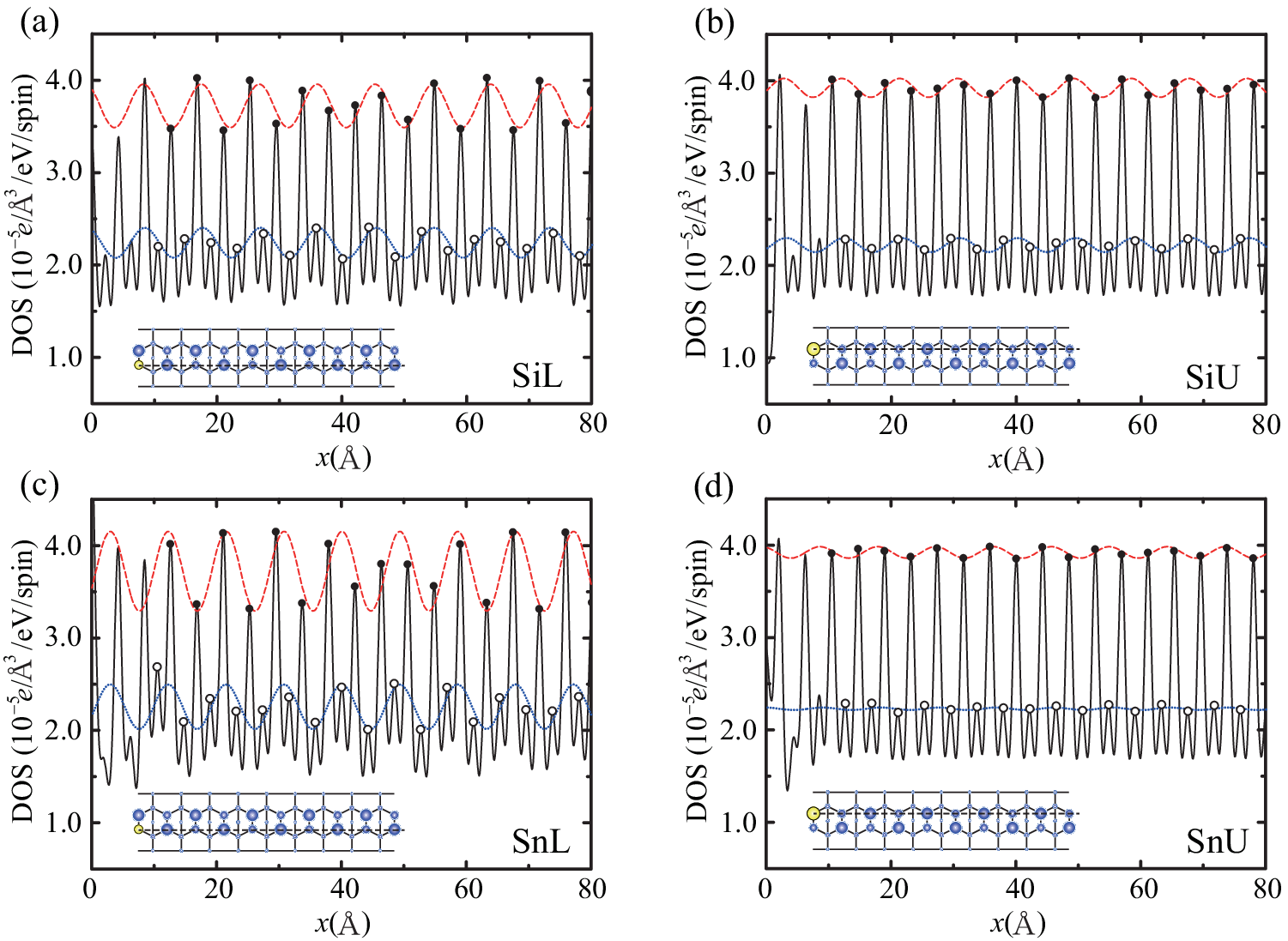}
\caption{Line profiles along dimer row including impurity atom indicated by dashed lines in insertions, where the meanings of symbols are the same as those in Fig.~\ref{fig:1}. Solid curve represents the standing wave. Dashed and dotted curves are fitted by Eq.~(\ref{eqn:fit}) using the densities above lower atoms of impurity side ($\bullet$) and upper atoms of impurity side ($\circ$), respectively. \label{fig:3}}
\end{center}
\end{figure*}

\begin{figure*}
\begin{center}
\includegraphics{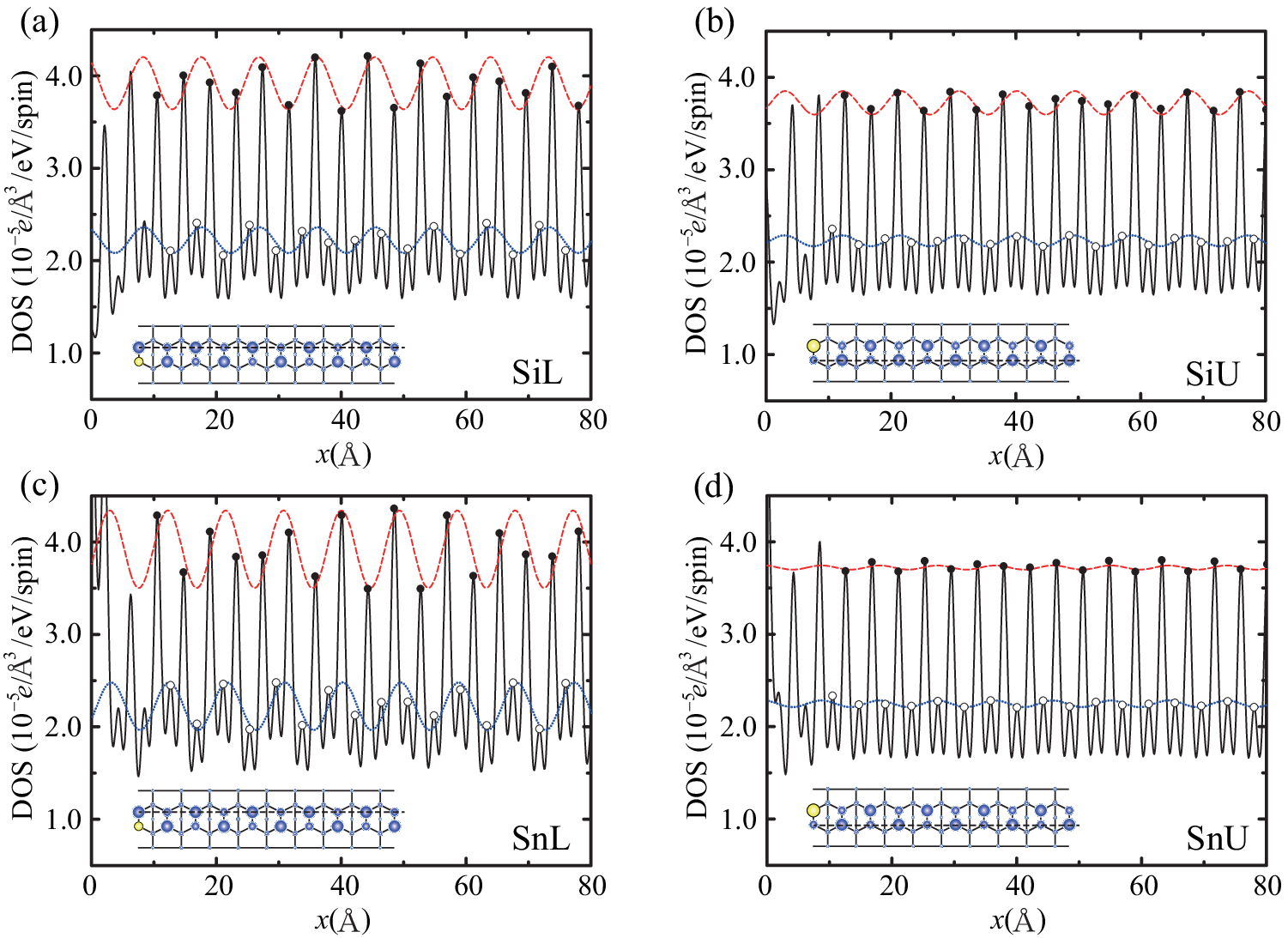}
\caption{Line profiles along dimer row not including impurity atom indicated by dashed lines in insertions, where the meanings of symbols are the same as those in Fig.~\ref{fig:1}. Solid curve represents the standing wave. Dashed and dotted curves are fitted by Eq.~(\ref{eqn:fit}) using the densities above lower atoms of nonimpurity side ($\bullet$) and upper atoms of nonimpurity side ($\circ$), respectively. \label{fig:4}}
\end{center}
\end{figure*}

The line profiles of the standing waves along the dimer row including and not including the impurity atom are depicted in Figs.~\ref{fig:3} and \ref{fig:4}, respectively. It should be noted that the charge density of the $\pi^*$ band above the lower atom of the dimer is larger than that above the upper atom. To demonstrate the period and phase shift of the standing waves clearly, the density of the standing waves is fitted by
\begin{equation}
\alpha(x)=A\cos(2k_x x+\phi),
\label{eqn:fit}
\end{equation}
where $A$ and $\phi$ are the amplitude and phase shift of the standing waves, respectively.\cite{comment} The fitting is carried out using the density of the standing waves above the lower (upper) atoms of the dimers, which are indicated by ``$\bullet$'' (``$\circ$'') in Figs.~\ref{fig:3} and \ref{fig:4}. Table~\ref{tbl:1} shows the amplitude and phase shift of the standing waves. The amplitude of the SnL (SiU) dimer is larger (smaller) than that of the SiL dimer, and the phase shifts of the SiU and SnL dimers are negative, while that of the SiL dimer is negative. These relationships do not change by the points employed for the fitting and these findings consistent with experimental results.\cite{tomatsu1} I have assured that the phase shift is not affected by the energy of the propagating waves significantly; the phase shifts for the energy of $E_F$+0.45 eV are listed in Table~\ref{tbl:2}, when they are computed using the density above the lower atoms including the impurities.

\begin{table*}
\caption{Amplitude and phase shift of the standing waves around the Ge-Si and Ge-Sn dimers, which are computed by fitting using densities above lower atoms of impurity side (``$\bullet$'' in Fig.~\ref{fig:3}), lower atoms of nonimpurity side (``$\bullet$'' in Fig.~\ref{fig:4}), upper atoms of impurity side (``$\circ$'' in Fig.~\ref{fig:3}), and lower atoms of nonimpurity side (``$\circ$'' in Fig.~\ref{fig:4}). The energy of the incident waves is $E_F$+0.55 eV.}
\begin{tabular}{lcccc}
\hline\hline 
Model & SiL & SiU & SnL & SnU \\ \hline
\multicolumn{5}{l}{Amplitude $A$ ($10^{-7}$ {\it e}/\AA$^3$/eV/spin)} \\
\: \: Lower atoms of impurity side    & $3.908$ & $1.692$ & $7.169$ & $1.031$ \\
\: \: Lower atoms of nonimpurity side & $4.706$ & $2.117$ & $6.985$ & $0.373$ \\
\: \: Upper atoms of impurity side    & $2.701$ & $1.275$ & $4.023$ & $0.219$ \\
\: \: Upper atoms of nonimpurity side & $2.320$ & $0.976$ & $4.289$ & $0.598$ \\
\multicolumn{5}{l}{Phase shift $\phi$ ($\pi$ rad)} \\
\: \: Lower atoms of impurity side    &  $0.221$ & $-0.602$ & $-0.650$ &  $0.142$ \\
\: \: Lower atoms of nonimpurity side &  $0.211$ & $-0.664$ & $-0.634$ &  $0.107$ \\
\: \: Upper atoms of impurity side    &  $0.199$ & $-0.678$ & $-0.642$ &  $0.056$ \\
\: \: Upper atoms of nonimpurity side &  $0.201$ & $-0.609$ & $-0.667$ &  $0.121$ \\
\hline\hline
\end{tabular}
\label{tbl:1}
\end{table*}

\begin{table*}
\caption{Phase shift of the standing waves at $E_F$+0.45 eV, which are computed by fitting using densities above lower atoms of impurity side (``$\bullet$'' in Fig.~\ref{fig:3}).}
\begin{tabular}{lcccc}
\hline\hline 
Model & SiL & SiU & SnL & SnU \\ \hline
Phase shift $\phi$ ($\pi$ rad) &  $0.258$ & $-0.542$ & $-0.554$ &  $0.085$ \\
\hline\hline
\end{tabular}
\label{tbl:2}
\end{table*}

\begin{table*}
\caption{Reflection coefficients obtained by OBM method, barrier heights, and barrier lengths calculated from the reflection coefficients using Eq.~(\ref{eqn:coefficient}). The energy of the incident waves is $E_F$+0.55 eV.}
\begin{tabular}{lcccc}
\hline\hline 
Model            & SiL & SiU & SnL & SnU \\ \hline
Coefficient      & $ 0.0519+0.0565i$ & $-0.0044-0.0270i$ & $-0.0293-0.1092i$ & $ 0.0138+0.0097i$ \\
Height $V$ (V)   & $-0.801$ & $ 4.444$ & $ 1.394$ & $-0.152$ \\
Length $a$ (\AA) &  $5.203$ &  $3.518$ &  $4.858$ &  $4.378$ \\ \hline\hline
\end{tabular}
\label{tbl:3}
\end{table*}

\begin{table}
\caption{The barrier heights for the incident waves with the energy of $E_F$+0.45 eV.}
\begin{tabular}{lcccc}
\hline\hline 
Model            & SiL & SiU & SnL & SnU \\ \hline
Height $V$ (V)   & $-0.509$ & $1.186$ & $0.275$ & $-0.201$ \\
\hline\hline
\end{tabular}
\label{tbl:4}
\end{table}

Let us discuss the scattering property of the defects in more detail. The OBM method provides the reflection coefficients of the scattering waves, which give us information on the scattering potential upon using a one-dimensional free-electron-like model. Table~\ref{tbl:3} shows the calculated reflection coefficients and the reflection probabilities. The barrier height $V$ and length $a$ of the scattering potential are determined by
\begin{equation}
c^{ref}=\frac{(k^2-K^2)(1-\e^{4iKa})\e^{2ika}}{(k+K)^2-(k-K)^2\e^{4iKa}},
\label{eqn:coefficient}
\end{equation}
which is derived from the penetration of electrons into a one-dimensional square potential barrier on the basis of quantum mechanics. Here, $k=v/\hbar$ and $K=\sqrt{(v^2-2mV)}/\hbar$, where $\hbar$ is the reduced Planck's constant, $m$ is the electron mass, and $v$ is the group velocity of the incident wave. $V$ and $a$ are fit so that the reflection coefficients correspond to those obtained by the OBM method. The calculated $V$ and $a$ are shown in Table II. The SiU and SnL dimers behave as a potential barrier, whereas the SiL and SnU dimers behave as a potential well. As shown in Table~\ref{tbl:4}, the energy of the scattering waves does not affect the sign of the scattering potential strongly. The signs of the scattering potential height except for that of the SnU dimer agree with those reported by Tomatsu {\it et al.},\cite{tomatsu1} where the potentials are estimated by fitting the line profile of the STS spectrum and/or by DFT calculation using periodic supercells. However, the potential of the SnU dimer, which is not observed in the experiment, is not in agreement. The computational model in Ref.~\onlinecite{tomatsu1} is under the periodic boundary conditions, in which the surface including the impurity atom is repeated continuously, while the surface with the impurity atom is suspended between the surfaces without impurities in the present study, which might result in the difference in the sign of the scattering potentials.

\begin{figure*}
\begin{center}
\includegraphics{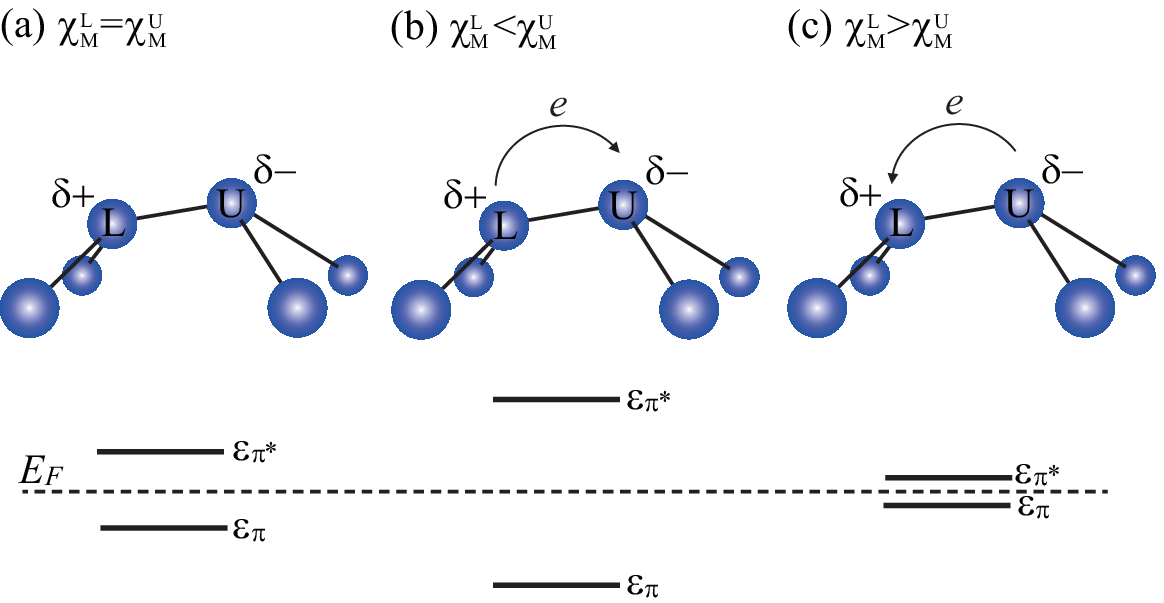}
\caption{Schematic image for opening the energy gap between $\varepsilon_\pi$ and  $\varepsilon_{\pi^*}$. $\chi_M^U$ ($\chi_M^L$) indicates the Mulliken electronegativity of upper (lower) atom and $\delta -$ ($\delta +$) is excess (lack) of electron charge due to the buckling of the dimer. \label{fig:5}}
\end{center}
\end{figure*} 

It is known that an electron conducts through the $\pi^*$ band of the Ge-Ge dimer on the Ge(001) surface. According to the explanation by Haneman,\cite{haneman} the $sp^3$ bonding state is stabilized as the bond angle around the atom of the dimer decreases. In the case of a Ge(001) surface, the $\pi$ band accumulating around the upper atom of the dimer is occupied after the surface reconstruction, as shown in Fig.~\ref{fig:5}(a). When the electronegativity of the upper atom is larger than that of the lower atom, the energy of the $\pi$ ($\pi^*$) band further decreases (increases) owing to the low (high) electronegativity of the lower (upper) atom [Figs.~\ref{fig:5}(b) and \ref{fig:5}(c)]. Since the Mulliken electronegativities of Si, Ge, and Sn are 2.0, 1.9, and 1.8, respectively, the energy of the $\pi^*$ bands of the SiU and SnL dimers is higher than that of the SiL and SnU dimers [Figs.~\ref{fig:5}(b) and \ref{fig:5}(c)]. Thus, the scattering potential of the SiU and SnL dimers acts as a barrier for the conducting electrons, whereas that of the SiL and SnU dimers acts as a well. In addition, because the electrons conduct through the states accumulating at the lower atom of the dimer, the length of the scattering potential barrier increases long when the impurity atom exists at the lower side of the dimer. In Ref.~\onlinecite{tomatsu1}, the projected charge density distribution of the $\pi$ band of the SiL dimer is high at the defect, while it is low in the case of the SnL dimer. Moreover, the energy gap between the $\pi$ and $\pi^*$ bands, $\varepsilon_{\pi^*}-\varepsilon_\pi$, of the SnL dimer is larger than that of the SiL dimer, which supports the explanation.

\section{Summary}
\label{sec:Summary}
The scattering potential of the defects on Ge(001) surface is investigated by first-principles calculation. The phase shifts of the standing waves due to the defects are in agreement with those obtained by experiments.\cite{tomatsu1} By calculating the reflection coefficients of the scattering wave functions, it was found that the scattering potentials of the SiL and SnU dimers act as a well, while those of the SiU and SnL dimers behave as a barrier. This characteristic is interpreted in terms of the electronegativity of the defects; when the electronegativity of the upper site of the dimer is large, the energy gap between the $\pi$ and $\pi^*$ bands increases, resulting in the generation of the potential barrier for the $\pi^*$ electrons. This explanation is also consistent with the LDOS in the vicinity of the dimers and with the value of $\varepsilon_{\pi^*}-\varepsilon_\pi$ computed using a conventional periodic supercell. In addition, the results demonstrate the applicability of first-principles transport calculations to the investigation of the scattering potential and phase shift due to defects on surfaces in collaboration with STM and STS measurements.

\section*{Acknowledgements}
The author would like to thank Prof. Fumio Komori and Dr. Kota Tomatsu of the University of Tokyo for providing the geometries of the Ge surfaces and for useful discussions. In addition, the author is indebted to Prof. Yoshitada Morikawa and Mr. Shoichiro Saito of Osaka University and Prof. Osamu Sugino of the University of Tokyo for fruitful discussions. This research was partially supported by Strategic Japanese-German Cooperative Program from Japan Science and Technology Agency and Deutsche Forschungsgemeinschaft, by the Computational Materials Science Initiative (CMSI), and by a Grant-in-Aid for Scientific Research on Innovative Areas (Grant No. 22104007) from the Ministry of Education, Culture, Sports, Science and Technology, Japan. The numerical calculation was carried out using the computer facilities of the Institute for Solid State Physics at the University of Tokyo and Center for Computational Sciences at University of Tsukuba.


\begin{references}
\bibitem{comment}
Since the present calculation are carried out at zero temperature, the decay factor for the energy relaxation of Eq. (1) in Ref.~\onlinecite{tomatsu1} is omitted.
\end{references}
\end{document}